\documentclass[preprint]{revtex4}

\begin{document}

\title{ Boltzmann-Gibbs Distribution of Fortune and
    Broken Time-Reversible Symmetry in Econodynamics}

\author{ P. Ao }

\address{ Department of Mechanical Engineering,
          University of Washington, Seattle, WA 98195, USA }

\date{June 10, (2005)}

\begin{abstract}
 Within the description of stochastic differential equations it is argued
 that the existence of Boltzmann-Gibbs type distribution in economy
 is independent of the time reversal symmetry in econodynamics.
 Both power law and exponential distributions can be accommodated
 by it.
 The demonstration is based on a mathematical structure discovered
 during a study in gene regulatory network dynamics.
 Further possible analogy between equilibrium economy
 and thermodynamics is explored.  \\
 { } \\
 PACS numbers: \\
89.65.Gh  Economics; econophysics, financial markets, business and
management; \\
05.10.Gg  Stochastic analysis methods   (Fokker-Planck, Langevin,
etc) \\
05.20.-y  Classical statistical mechanics.  \\
87.23.Ge  Dynamics of social systems

\end{abstract}


\maketitle

\section{Introduction}

Carefully and extensive analysis of real economic and financial
data have revealed various exponential and power-law distributions
regarding to money, income, wealth, and other economic quantities
in ancient and modern social societies \cite{ms,bp,dy,bm}. A
remarkable analogy between the economic system and a
thermodynamical system has been revealed in a recent study
\cite{dy}. Using a detailed microdynamical model with time
reversal symmetry, it was demonstrated that a Boltzmann-Gibbs type
distribution exists in economic systems. Indeed, ample empirical
analysis support such suggestion \cite{dy,sy}. Nevertheless,
different microdynamical models lead to apparently different
distributions \cite{bm,dy}. Those distributions are supported by
empirical data, too. The nature of such difference may reveal the
difference in corresponding economic structure. For example, this
difference has been tentatively attributed to the role played by
time reversal symmetry in microdynamical models \cite{dy}.

Following the tradition of synthesizing interdisciplinary
knowledge, such as from biology, physics, and finance
\cite{farmer}, in this letter we argue that irrespective of the
time reversal symmetry the Boltzmann-Gibbs type distribution
always exists. A broader theoretical base is thus provided. The
demonstration is performed within the framework of stochastic
differential equations and is based on a novel mathematical
structure discovered during a recent study in gene regulatory
network dynamics \cite{zhu}. In the light of thermodynamical
description, possible explanations for the origin of the
difference in various empirical distributions are proposed.

\section{Boltzmann-Gibbs distribution in finance}

Stochastic differential equations or Langevin equations and their
corresponding Fokker-Planck equations have been shown to be a
useful modelling tool in economy \cite{bs,heston,ms,bp}. One of
the best examples is the Black-Scholes formula in option pricing
theory \cite{bs}. This kind of mathematical formulation provides a
direct connections between the microdynamics and the stationary
state and has been used to generate various distribution laws
\cite{ms,bp,bm,dy,sy}.

Specifically, the stochastic differential equation may take the
following form:
\begin{equation}  \label{standard}
  \dot{{\bf q}} = {\bf f}({\bf q}) + N_I({\bf q}) \xi(t) \; ,
\end{equation}
where ${\bf f}$ and ${\bf q}$ are $n$-dimensional vectors and
${\bf f}$ a nonlinear function of the state variable ${\bf q}$.
The state variable is the quantity to specify the economic system.
It may be the money, the income, or other suitable indices. The
noise ${\bf \xi}$ is a standard Gaussian white noise with $l$
independent components: $\langle \xi_i \rangle = 0$, $\langle
\xi_i(t) \xi_j (t')\rangle= 2 \delta_{ij} \delta (t-t')$, and
$i,j=1, 2, ..., l$. In Eq.(1) we explicitly factorize out the pure
noise source and the state variable dependent part for the
convenient of later description.

The specification of the noise in Eq.(\ref{standard}) is through
the $n\times n$ diffusion matrix $D({\bf q})$ by the following
matrix equation
\begin{equation}
  N_I({\bf q}) N_I^\tau ({\bf q}) = \epsilon \; D({\bf q}) \; ,
\end{equation}
where $N_I$ is an $n\times l$ matrix,  $N_I^\tau$ is its the
transpose, and $\epsilon$ a nonnegative numerical constant to keep
tract of the noise. It plays the role of temperature in
thermodynamics. According to Eq.(2) the $n\times n$ diffusion
matrix $D$ is both symmetric and nonnegative. For the dynamics of
state variable ${\bf q}$, all what needed from the noise is the
diffusion matrix $D$. Hence, it is not necessary to require the
dimension of the noise vector $\xi$ be the same as that of the
state vector ${\bf q}$ and to require more specific knowledge of
$n\times l$ matrices $\{ N_I \}$ beyond Eq.(2).

It is known that even in situations that Eq.(\ref{standard}) is
not an exact description, and perhaps it would never be in a
rigorous sense in economy, it may still serve as the first
approximation for further modelling \cite{ms,bp}. Indeed, it has
been empirically verified to be a rather accurate description in
economy \cite{bs,heston,rs,loffredo}. Because the energy function
or Hamiltonian has played a dominant role in equilibrium physics
processes, the crucial question is whether or not a similar
quantity exists in a more general setting. In the following we
present an argument leading to the positive answer.

There exists several ways to deal with the stochastic equations
equation in the form of Eq.(1) and (2). The most commonly used are
those of Ito and Stratonovich methods \cite{vankampen,ms,bp}.
However, with those methods the connection between the existence
of energy function like quantity in Eq.(1) and the stationary
distribution is not clear when the time reversal symmetry is
broken \cite{vankampen}. The difficulty for finding such potential
function can be illustrated by the fact that usually $D^{-1}({\bf
q}) {\bf f}({\bf q})$ cannot be written as the gradient of a
scalar function \cite{vankampen} in the absence of detailed
balance condition or in the broken time reversal symmetry. This
will become precise as we proceed.

During a recent study of the robustness of the genetic switch in a
living organism \cite{zhu}, it was discovered that Eq.(1) can be
transformed into the following form,
\begin{equation} \label{normal}
  [ A({\bf q}) + C({\bf q})] \dot{{\bf q}} = \partial_{\bf q}
   \phi({\bf q}) + N_{II}({\bf q})\xi(t) \; ,
\end{equation}
where the noise $\xi(t)$ is from the same source as that in
Eq.(1). Here we tentatively name $A({\bf q})$ the adaptation
matrix, $C({\bf q})$ the conservation matrix, and the scalar
function $\phi({\bf q})$ the fortune function. The gradient
operation in state space is denoted by $\partial_{\bf q}$. The
adaptation matrix $A({\bf q})$ is defined through the following
matrix equation
\begin{equation}
  N_{II}({\bf q}) N_{II}^\tau ({\bf q})=  \epsilon \; A({\bf q}) \; ,
\end{equation}
which guarantees that $A$ is both symmetric and nonnegative. The
$n\times n$ conservation matrix $C$ is antisymmetric. We define
$$
  A({\bf q}) + C({\bf q})
  = 1/[ D({\bf q}) + Q({\bf q})] \equiv M({\bf q}) \; .
$$
with the $n\times n$ matrix $M$ is the solution of following two
matrix equations \cite{ao04}. The first equation is the potential
condition
\begin{equation}
 \partial_{\bf q} \times [ M({\bf q}) {\bf f}({\bf q}) ] = 0 \; ,
\end{equation}
which gives $n(n-1)/2$ conditions
 [the wedge product for two arbitrary vectors ${\bf v}_1$ and ${\bf v}_2$ in n-dimension:
$ [{\bf v}_1 \times
   {\bf v}_{2}]_{ij} = {v_1}_i {v_2}_j - {v_1}_j {v_2}_i
  \; , \; i,j = 1,2, ... , n $ ].
The second equation is the generalized Einstein relation between
the adaptation and diffusion matrices in the presence of
conservation matrix
\begin{equation}
 M({\bf q}) D({\bf q}) M^{\tau}({\bf q})
  = \frac{1}{2} [M({\bf q}) + M^{\tau}({\bf q}) ] \; .
\end{equation}
which gives $n(n+1)/2$ conditions \cite{ao04}.
 The fortune
function $\phi({\bf q})$ is connected to the deterministic force
${\bf f}({\bf q})$ by
$$
  \partial_{\bf q} \phi({\bf q}) = M({\bf q}) {\bf f}({\bf q}) \; .
$$
For simplicity we will assume $\det(A) \neq 0$ in the rest of the
letter. Hence $\det(M) \neq 0$ \cite{kat}. Thus, the adaptation
matrix $A$, the conservation matrix $Q$ and the fortune function
$\phi$ in Eq.(3) and (4) can be completely determined by Eq.(1)
and (2). The breakdown of detailed balance condition or the time
reversal symmetry is represented by the finiteness of the
conservation matrix
\begin{equation}
  C({\bf q}) \neq 0  \; ,
\end{equation}
or equivalently $Q \neq 0$. The usefulness of the formulation of
Eq.(3) and (4) is already manifested in the successful solution of
outstanding stable puzzle in gene regulatory dynamics \cite{zhu}
and in solving two fundamental controversies in population
genetics \cite{ao05}.

A few remarks on Eq.(3) are in order. In the light of classical
mechanics in physics, Eq.(3) is in precisely the form of Langevin
equation. The fortune function $\phi$ corresponds to the potential
function but opposite in sign to reflect the fact that in economy
there is a tendency to seek the peak or maximum of fortune. The
adaptive matrix $A$ plays the role of friction. It represents
adaptive dynamics and is the dynamical mechanism to  seek the
nearby fortune peak. The conservation matrix $C$ plays the role
analogous to a magnetic field. Its dynamics is similar to that of
the Lorentz force, hence conserves the fortune. As in classical
mechanics, the finiteness of the conservation matrix $C$ breaks
the time reversal symmetry.

It was heuristically argued \cite{ao04} and rigorous demonstrated
\cite{yin} that the stationary distribution $\rho({\bf q})$ in the
state space is, if exists,
\begin{equation} \label{bg-dis}
 \rho({\bf q}) \propto  \exp\left( {\phi({\bf q})
   \over{\epsilon} } \right) \; .
\end{equation}
Therefore, the fortune function $\phi$ acquires both the dynamical
meaning through Eq.(\ref{normal}) and the steady state meaning
through Eq.(\ref{bg-dis}). Specifically, in the so-called
zero-mass limit to differentiate from Ito and Stratonovich
methods, the Fokker-Planck equation for the probability
distribution $\rho({\bf q},t)$ takes the form \cite{yin}
\begin{equation} \label{fp-eq}
 {\partial_t \rho({\bf q},t) }
  = \partial_{{\bf q}}^{\tau} M^{-1}({\bf q}) [\epsilon \partial_{\bf q}
   - \partial_{{\bf q}} \phi({\bf q})] \rho({\bf q},t) \; .
\end{equation}
Here $\partial_t$ is a derivative with respect to time and
$\partial_{\bf q}$ represents the gradient operation in state
space. We note that Eq.(\ref{bg-dis}) is a stationary solution to
Eq.(\ref{fp-eq}) even it may not be normalizable, that is, even
when the partition function ${\cal Z} = \int d^{n}{\bf q} \;
\rho({\bf q})$ is ill-defined. Again, we emphasize that no time
reversal symmetry is assumed in reaching this result. This
completes our demonstration on the existence of the
Boltzmann-Gibbs distribution in economy.

Using $ M^{-1}({\bf q})= D({\bf q}) + Q({\bf q})$ and $ {\bf
f}({\bf q}) = [D({\bf q}) + Q({\bf q})] \partial_{{\bf q}}
\phi({\bf q})$, Eq.(\ref{fp-eq}) can be rewritten in a more
suggestive form \cite{yin}
\begin{equation}  \label{fp-eq2}
 {\partial_t \rho({\bf q},t) }
  = \partial_{{\bf q}}^{\tau} [\epsilon D({\bf q})\partial_{\bf q}
   + \epsilon (\partial_{{\bf q}}^{\tau} Q({\bf q}))
   -  {\bf f}({\bf q})] \rho({\bf q},t) \; .
\end{equation}
It is clear that in the presence of time reversal symmetry, {\it
i.e.} $Q = 0$, one can directly read the fortune function $\phi$
from above form of Fokker-Planck equation as $
\partial_{{\bf q}} \phi({\bf q}) = D^{-1}({\bf q}) {\bf f}({\bf
q})$.

For the sake of completeness, we list the Fokker-Placnk equations
corresponding to Ito and Stratonovich treatments of Eq.(1) and (2)
\cite{vankampen}:
\begin{equation} \label{ito}
  \partial_t \rho_{I}({\bf q},t)
   = \sum_{i=1}^{n}\partial_{ q_i} \left[ - f_i({\bf q})
     + \epsilon \sum_{j=1}^n \partial_{ q_j} D_{i,j}({\bf q}) \right]
       \rho_{I}({\bf q}, t) \; ,  {\ } {\ } {\ } (Ito)
\end{equation}
and
\begin{equation} \label{str}
  \partial_t \rho_{S}({\bf q},t)
   = \sum_{i=1}^n \partial_{ q_i} \left[ - f_i({\bf q})
     + \sum_{j=1}^n\sum_{k=1}^l {N_{I}}_{ik}({\bf q})
           \partial_{ q_j} {N_I}_{jk}({\bf q}) \right]
             \rho_S({\bf q}, t) \; .  {\ } {\ } {\ } (Str)
\end{equation}
The connection of fortune function with both dynamical (Eq.(1) and
(2)) and stationary state is indeed not clear in above two
equations. Nevertheless, it has been shown \cite{yin} that there
are corresponding fortune functions, adaptation and conservation
matrices to Eq.(\ref{ito}) and (\ref{str}). We point out here that
when the matrix $N_I$ is independent of state variable,
Eq.(\ref{ito}) and (\ref{str}) are the same but may still differ
from Eq.(\ref{fp-eq}), because the gradient of the antisymmetric
matrix $Q({\bf q})$ may not be zero. This last property shows that
the time reversal symmetry is indeed important.

\section{Two Examples}

The Fokker-Planck equation used by Silva and Yakovenko \cite{sy}
has the form:
\begin{equation} \label{sy-eq}
 {\partial_t \rho_{sy}({ q},t) }
  = \partial_{ q} [
    a({ q}) +  \partial_{ q} b({ q})] \rho_{sy}({ q},t) \; ,
\end{equation}
and Fokker-Planck equation used by Bouchaud and Mezard \cite{bm}
has the form
\begin{equation} \label{bm-eq}
 {\partial_t \rho_{bm}({ q},t) }
  = \partial_{ q} [(J (q-1) + \sigma^2 q
   + \sigma^2 q \partial_{ q} q  ] \rho_{bm}({ q},t) \; .
\end{equation}
They are all in one dimension. We immediately conclude that the
conservation matrix $C$, equivalently $Q$, is zero, because there
is no conservation matrix in one dimension. In accordance with the
definition in the present letter, which is consistent with that in
nonequilibrium processes \cite{vankampen}, the dynamics described
by above two equations can be effectively classified as time
reversal symmetric.

Rewriting them in symmetric form with respect to the derivative of
state variable $q$ as in Eq.(\ref{fp-eq2}), we have
\begin{equation} \label{sy-eq}
 {\partial_t \rho_{sy}({ q},t) }
  = \partial_{ q} [ a({ q}) + (\partial_{ q} b({ q}))
     + b({ q}) \partial_{ q}] \rho_{sy}({ q},t) \; ,
\end{equation}
and
\begin{equation} \label{bm-eq}
 {\partial_t \rho_{bm}({ q},t) }
  = \partial_{ q} [(J (q-1) + \sigma^2 q
   + \sigma^2 q +  \sigma^2 q \partial_{ q}  ] \rho_{bm}({ q},t) \; .
\end{equation}
The corresponding wealth functions can be immediate read out as
\begin{eqnarray}
 \phi_{sy}({ q})
  & = & - \int_{q_0}^q d q'
   { a(q') + (\partial_{q'} b(q')) \over{ b(q') } }  \nonumber \\
  & = & - \frac{a}{b} (q-q_0)  \; , \; if \;  a, b = constant  \\
 \phi_{bm}({ q})
  & = & - \int_{q_0}^{q} d q'
    {(J (q'-1) + 2 \sigma^2 q' \over { \sigma^2 {q'}^2 } }  \nonumber \\
  & = & - {J \over{\sigma^2 }}  {1 \over{q}}
        - \left( 2 + {J \over{\sigma^2}} \right) \ln q
        + {J \over{\sigma^2 }}  {1 \over{q_0}}
        + \left( 2 + {J \over{ \sigma^2}} \right) \ln q_0
\end{eqnarray}
They are exactly what found in Ref.[\cite{sy}] and [\cite{bm}]:
the first one corresponds to an exponential distribution and the
second one a power law distribution according to the
Boltzmann-Gibbs distribution Eq.(\ref{bg-dis}).

\section{Ensembles and state variables}

Having defined a precise meaning of time reversal symmetric and
have demonstrated that the Boltzmann-Gibbs distribution even in
the absence of time reversal symmetry, we explore further
connection between econodynamics and statistical physics.

There are two general types of situations which would generate
different distributions in statistical physics and thermodynamics.
The first one is to link to constraints on the system under
various conditions. In statistical physics such constraints are
described by various ensembles and free energies. For example,
there are canonical and grand-canonical ensembles. There are Gibbs
and Helmholtz free energies, entropy, enthalpy, etc. Those
ensembles have their characteristic distributions. It would be
interesting to know the corresponding situations in economy.

Even with a given constraint, the form of distribution depends on
the choice of state variable.  For example, for ideal gas model,
the distributions are different if views from the kinetic energy
and from velocity. Hence, there is a question of appropriate state
variable for a given situation, with which the physics becomes
particular transparent. It would be interesting to know what be
the appropriate variables to describe an economic system. Within
this context, the difference between what discovered by Dragulecu
and Yakovenko \cite{dy} and Bouchaud and Mezard \cite{bm} is
perhaps more due to the difference in choices of state variables,
because it seems they are describing the same situation of same
system under same constraints.

A remark on terminology is in order. It was demonstrated above
that regardless of the time reversal symmetry the Boltzmann-Gibbs
distribution, Eq.(\ref{bg-dis}), exists. The fortune function
$\phi$ has an additional dynamical meaning defined in
Eq.(\ref{normal}). Both exponential and power law distribution can
be represented by Eq.(\ref{bg-dis}). In fact, it is well known
that power law distributions exist in statistical physics. A
nontrivial example is the Kosterlitz-Thouless transition
\cite{kt}.  Thus it does not appear appropriate to call the power
law distribution non-Boltzmann-Gibbs distribution. Such a
terminology confusion was already noticed before \cite{ls}.

In the view of the dominant role of entropy in Kosterlitz-Thouless
transition \cite{kt}, the ubiquitous existence of power law
distribution in economy may suggest that the entropy effect is
rather important in econodynamics. This may corroborate with the
suggestion of ``superthermal'' in economy \cite{sy}.

\section{Conclusions}

In this letter we demonstrate that the existence of
Botlzmann-Gibbs distribution in finance is independent of time
reversal symmetry. Both power law and exponential distributions
are within its description. In analogous to similar situation in
statistical physics, the differences among those distributions
discovered empirically in economy are likely the result of
different choices of state variables to describe the same system
in econodynamics.

{\ }

This work was supported in part by USA NIH grant under HG002894.

\end{document}